\documentclass[twoside,twocolumn,9pt]{article}
\usepackage{extsizes}
\usepackage[super,sort&compress,comma]{natbib} 
\usepackage[version=3]{mhchem}
\usepackage[left=1.5cm, right=1.5cm, top=1.785cm, bottom=2.0cm]{geometry}
\usepackage{balance}
\usepackage{mathptmx}
\usepackage{sectsty}
\usepackage{graphicx} 
\usepackage{lastpage}
\usepackage[format=plain,justification=justified,singlelinecheck=false,font={stretch=1.125,small,sf},labelfont=bf,labelsep=space]{caption}
\usepackage{float}
\usepackage{fancyhdr}
\usepackage{fnpos}
\usepackage[english]{babel}
\addto{\captionsenglish}{%
  
}
\usepackage{array}
\usepackage{droidsans}
\usepackage{charter}
\usepackage[T1]{fontenc}
\usepackage[usenames,dvipsnames]{xcolor}
\usepackage{setspace}
\usepackage[compact]{titlesec}
\usepackage{hyperref}

\usepackage{epstopdf}

\definecolor{cream}{RGB}{222,217,201}

\begin{document}

\pagestyle{fancy}
\thispagestyle{plain}
\fancypagestyle{plain}{
\renewcommand{\headrulewidth}{0pt}
}

\makeFNbottom
\makeatletter
\renewcommand\LARGE{\@setfontsize\LARGE{15pt}{17}}
\renewcommand\Large{\@setfontsize\Large{12pt}{14}}
\renewcommand\large{\@setfontsize\large{10pt}{12}}
\renewcommand\footnotesize{\@setfontsize\footnotesize{7pt}{10}}
\makeatother

\renewcommand{\thefootnote}{\fnsymbol{footnote}}
\renewcommand\footnoterule{\vspace*{1pt}%
\color{cream}\hrule width 3.5in height 0.4pt \color{black}\vspace*{5pt}} 
\setcounter{secnumdepth}{5}

\makeatletter 
\renewcommand\@biblabel[1]{#1}            
\renewcommand\@makefntext[1]%
{\noindent\makebox[0pt][r]{\@thefnmark\,}#1}
\makeatother 
\renewcommand{\figurename}{\small{Fig.}~}
\sectionfont{\sffamily\Large}
\subsectionfont{\normalsize}
\subsubsectionfont{\bf}
\setstretch{1.125} 
\setlength{\skip\footins}{0.8cm}
\setlength{\footnotesep}{0.25cm}
\setlength{\jot}{10pt}
\titlespacing*{\section}{0pt}{4pt}{4pt}
\titlespacing*{\subsection}{0pt}{15pt}{1pt}

\fancyfoot{}
\fancyfoot[LO,RE]{\vspace{-7.1pt}\includegraphics[height=9pt]{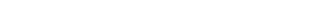}}
\fancyfoot[CO]{\vspace{-7.1pt}\hspace{13.2cm}\includegraphics{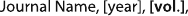}}
\fancyfoot[CE]{\vspace{-7.2pt}\hspace{-14.2cm}\includegraphics{RF}}
\fancyfoot[RO]{\footnotesize{\sffamily{1--\pageref{LastPage} ~\textbar  \hspace{2pt}\thepage}}}
\fancyfoot[LE]{\footnotesize{\sffamily{\thepage~\textbar\hspace{3.45cm} 1--\pageref{LastPage}}}}
\fancyhead{}
\renewcommand{\headrulewidth}{0pt} 
\renewcommand{\footrulewidth}{0pt}
\setlength{\arrayrulewidth}{1pt}
\setlength{\columnsep}{6.5mm}
\setlength\bibsep{1pt}

\makeatletter 
\newlength{\figrulesep} 
\setlength{\figrulesep}{0.5\textfloatsep} 

\newcommand{\topfigrule}{\vspace*{-1pt}%
\noindent{\color{cream}\rule[-\figrulesep]{\columnwidth}{1.5pt}} }

\newcommand{\botfigrule}{\vspace*{-2pt}%
\noindent{\color{cream}\rule[\figrulesep]{\columnwidth}{1.5pt}} }

\newcommand{\dblfigrule}{\vspace*{-1pt}%
\noindent{\color{cream}\rule[-\figrulesep]{\textwidth}{1.5pt}} }

\makeatother

\twocolumn[
  \begin{@twocolumnfalse}
{\includegraphics[height=30pt]{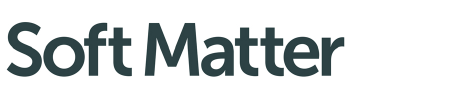}\hfill\raisebox{0pt}[0pt][0pt]{\includegraphics[height=55pt]{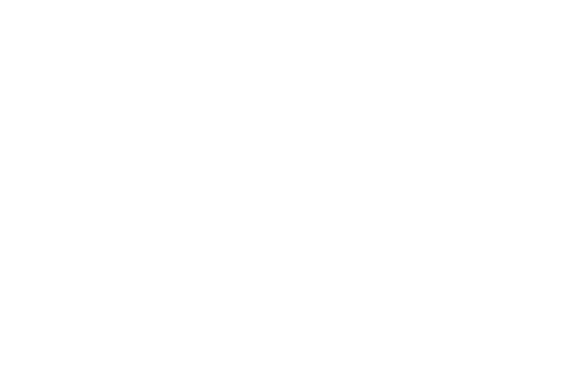}}\\[1ex]
\includegraphics[width=18.5cm]{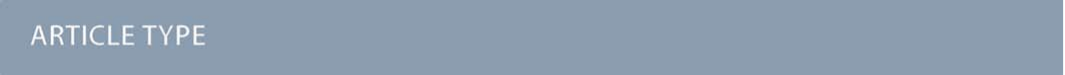}}\par
\vspace{1em}
\sffamily
\begin{tabular}{m{4.5cm} p{13.5cm} }

\includegraphics{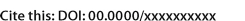} & 
\noindent\LARGE{\textbf{Understanding Polymer-Colloid Gels: A Solvent Perspective Using Low-Field NMR}} \\
\vspace{0.3cm} & \vspace{0.3cm} \\

 & \noindent\large{Léo Hervéou,\textit{$^{a,b,\mathsection}$} Gauthier Legrand,\textit{$^{b,\mathsection}$} Thibaut Divoux\textit{$^{b}$} and Guilhem P. Baeza$^{\ast,}$\textit{$^{a,\ddag}$}} \\


\includegraphics{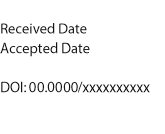} & \noindent\normalsize{
The present work emphasizes the relevance of low-field NMR relaxometry to investigate colloid-polymer hydrogels by probing water dynamics across a wide range of formulations between $\rm 10^{\circ}C$ and $\rm 80^{\circ}C$. By examining the temperature dependence of the transverse relaxation time $T_2$, we demonstrate a clear link between the NMR response and the rheological behavior of the hydrogels. In particular, we show that NMR relaxometry targeting the solvent provides reliable insights into the hydrogel microstructure and allows the detection of phase transitions and aging processes. Our findings suggest that this solvent-focused technique could greatly benefit the soft matter community, complementing other experimental methods in the study of gels. 
}  \\

\end{tabular}

 \end{@twocolumnfalse} \vspace{0.6cm}

  ]

\renewcommand*\rmdefault{bch}\normalfont\upshape
\rmfamily
\section*{}
\vspace{-1cm}


\footnotetext{\textit{$^{a}$~INSA Lyon, UCBL, CNRS, MATEIS, UMR5510, 69621, Villeurbanne, France; E-mail: guilhem.baeza@insa-lyon.fr}}

\footnotetext{\textit{$^{b}$~ENSL, CNRS, Laboratoire de physique, F-69342 Lyon, France }}


\footnotetext{\ddag~Present address: Univ. Jean Monnet, Ingénierie des Matériaux Polymères, UMR 5223, 20 rue Annino, 42000 Saint-Etienne}

\footnotetext{$\mathsection$~These authors contributed equally to this work}

\footnotetext{\dag~Supplementary Information available: 
1. $T_2=f(c_{\rm CMC})$ plots; 2. Cooling vs.~heating $T_2=f(1000/T)$ plots and $E_a$ values for the whole set of samples; 3. Corresponding viscous modulus of the elastic modulus plotted in Fig.~\ref{fgr:rheo}; 4. CPMG experiments with various echo times.}



Hydrogels represent a fascinating class of versatile materials originally consisting of a low fraction of hydrosoluble polymer forming a 3D percolated network in water.\cite{Appel:2012} Important progresses regarding their structural design have been made in the last decades,  notably leading to the emergence of double- \cite{gong2010double, chen2015fundamentals,Li:2024} and hybrid-networks,\cite{Messing:2011,li2014hybrid} respectively incorporating two (or more \cite{li2021high}) polymers and inorganic particles. These advanced materials exhibit enhanced mechanical properties, leading to widespread use in key sectors like the food industry,\cite{zhang2020applications,nath2022comprehensive} medical engineering,\cite{rose2014nanoparticle, ulijn2007bioresponsive,palmese2019hybrid} and energy.\cite{zhao2018nanostructured,shi2015nanostructured}  
Among them, natural polymer-based hydrogels relying on networks of proteins, polyesters, or polysaccharides present the advantage of employing abundant, biocompatible, and (sometimes) edible polymers. Their amphiphilic nature, coupled with well-chosen inorganic particles, results in a rich phase diagram where mechanical properties can be easily selected from the formulation and the processing conditions.\cite{Schexnailder:2009,Dehne:2017,Yu:2020} For example, we have recently demonstrated the dual nature of physical hydrogels made of the sodium salt of carboxymethylcellulose (CMC) and carbon black (CB) soot particles. Depending on their composition, CB-CMC hydrogels can either display a microstructure akin to a  \textit{colloidal gel} in which a percolated network of CB particles is stabilized by CMC, or a \textit{polymer gel} in which a CMC matrix is physically cross-linked by CB particles.\cite{legrand2023dual} 

Besides, the amphiphilic character of natural polymer-based hydrogels makes them particularly thermo-sensitive, including within the narrow temperature window defined by water crystallization and vaporization ($\rm 0$ and $\rm 100^{\circ}C$).\cite{moakes2015preparation} In addition to irreversible chemical alteration, temperature changes can significantly impact chain conformation (e.g., protein denaturation \cite{tanford1968protein}), or strongly influence the gelation scenario, leading to major topological changes of the network.\cite{fan2022thermosensitive} Such changes have been evidenced by multiple techniques, among which relaxometry, i.e., low-field NMR spectroscopy, was shown to yield precious insights regarding the microstructure of complex biological tissues\cite{Deoni:2011,Granziera:2015} and foodstuff.\cite{Salomonsen:2007,Harbourne:2011,Kirtil:2016,Ozel:2017} This technique, which is used to obtain information about the mobile protons of a sample, has also been recently coupled to rheometry offering time-resolved information on the samples microstructure under shear.\cite{Ratzsch:2017,Radebe:2020,Fengler:2022,Xiong:2023} 
In solid porous media, low-field NMR targets the solvent to compute the sample's pore size distributions.\cite{Maillet:2022} In contrast, in aqueous suspensions and hydrogels, where water is replaced by D$_2$O to mask the contribution from the proton of the solvent, low-field NMR targets the dynamics of the dispersed phase alone.

Here, we take a mixed approach by conducting Low Field (LF) $\rm ^1H$-NMR experiments on CB-CMC hydrogels to monitor the dynamics of {water}, i.e., the \textit{solvent}, over a wide range of compositions and temperatures. The use of the popular Carr-Purcell-Meiboom-Gill (CPMG) pulse routine\cite{Carr:1954,Meiboom:1958,Brown:2014} allows us to measure the spin-spin (transverse) relaxation time $T_2$, which, in the case of pure water is of several seconds.\cite{Fullerton:1982} 
Our results clearly show that LF-NMR experiments targeting the solvent can effectively detect the sol-gel transition, thermally-induced phase transition, and physical aging. This makes it a promising and complementary technique to the traditional focus on the dispersed phase.

\begin{figure}[!t]
\centering
  \includegraphics[width=1\columnwidth]{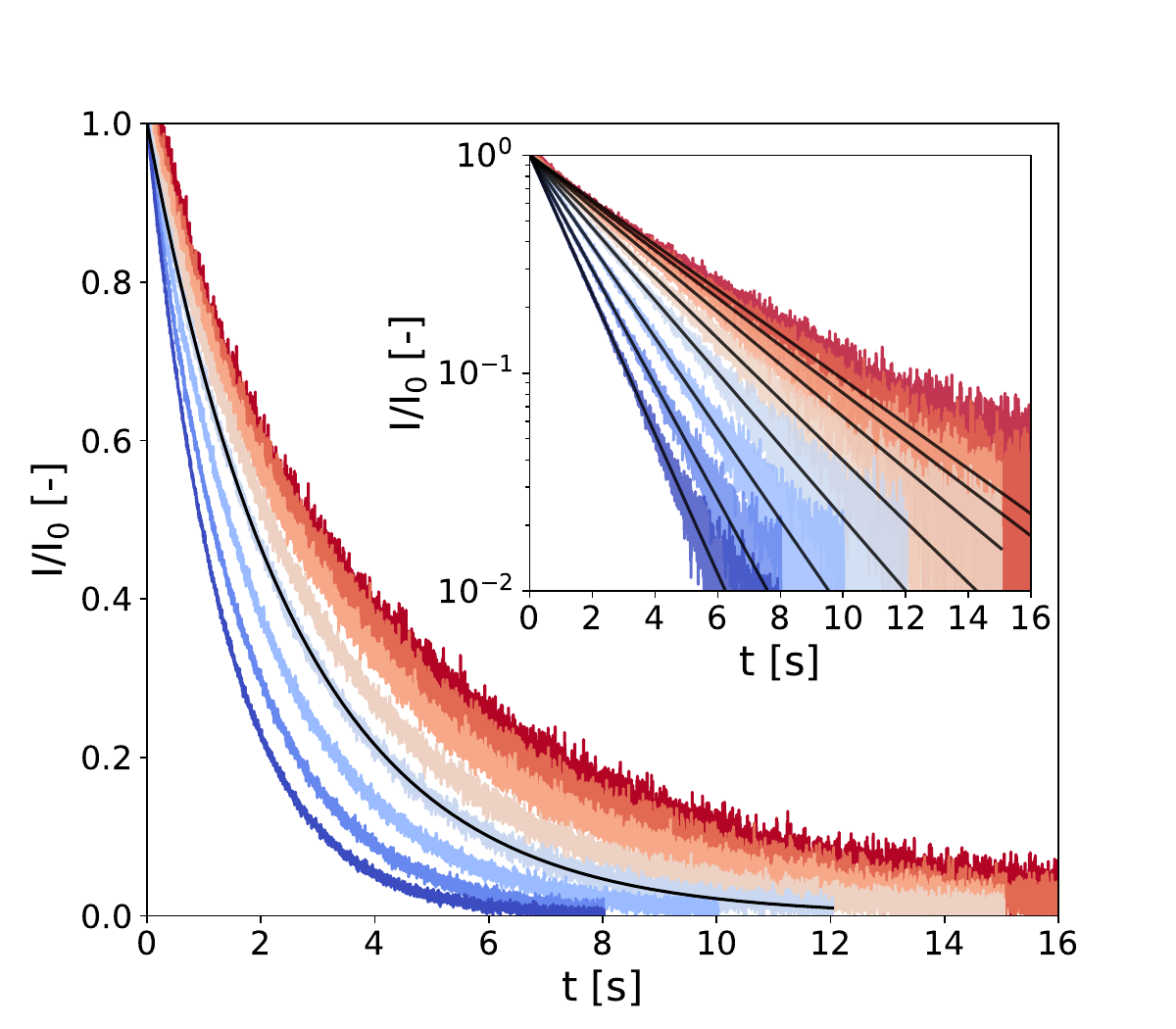}
  \caption{LF-NMR CPMG signal normalized by $I_0$, the signal amplitude extrapolated at t=0 s extracted from fitting the data with Eq.~\eqref{eq:1} (black solid lines). Measurements are performed on a CMC0.01-CB2 hydrogel at various temperatures, every $10~\rm ^{\circ}C$ between $10~\rm ^{\circ}C$ and $80~\rm ^{\circ}C$ (see color code from dark blue to red). Inset: same data in semi-logarithmic representation.}
  \label{fgr:I(t)}
\end{figure}

In the following, samples are denoted through the code "CMCX-CBY," where X and Y represent the mass fraction (in $\rm wt.\%$) of CB and CMC, respectively (see technical details in the \textit{experimental section}).
Typical CPMG experiment outputs are presented in Figure~\ref{fgr:I(t)} for a low-density gel, i.e., a hydrogel of composition CMC0.01-CB2 where the transverse magnetization of hydrogen atoms (almost exclusively coming from water) $I(t)$ decreases over a characteristic time $T_2$ following the expression 
\begin{equation} \label{eq:1}
    I(t)=I_0\ \exp(-t/T_2),
\end{equation}
where $I_0=I(0)$ and $t$ is the experimental time. 
As expected, $T_2$ increases for higher temperatures, indicating the overall greater mobility of water molecules. However, in contrast to several other natural polymer-based hydrogels,\cite{besghini2023time,capitani2001water,abrami2018use,Fengler:2022} the \textit{single} and \textit{unstretched} character of the exponential decay suggests that, on average, over the entire experimental time, all water molecules in the hydrogel exhibit identical dynamics. In other words, despite the spatial density fluctuations and the corresponding various residence times in different micro-environments, water molecules appear to move through the material much faster than the time required to lose their magnetization. This indicates the absence of significant molecular immobilization, which might have been expected in hydrophilic polymers.\cite{Ghi:2002,Kanekiyo:1998,Abrami:2014,Abrami:2023} A rough estimate of the diffusion length $L_D$, calculated using the self-diffusion coefficient of water at room temperature\cite{Wang:1965} ($D_{w}\approx 2.3\ 10^{-9} \rm m^2 s^{-1}$) and $T_2$ ($\approx 4~\rm s$), yields $L_D=\sqrt{D_{w}T_2}\approx 100~\rm \mu m$. This value is much larger than the mesh size of CMC gels determined from rheometry, which is approximately  $50~\rm nm$ \cite{legrand2024rheological} or the average diameter of CMC \textit{fringe micelles} measured by neutron scattering, approximately $75~\rm nm$.\cite{legrand2024acid}

\begin{figure*}[h]
 \centering
 \includegraphics[height=12.0cm]{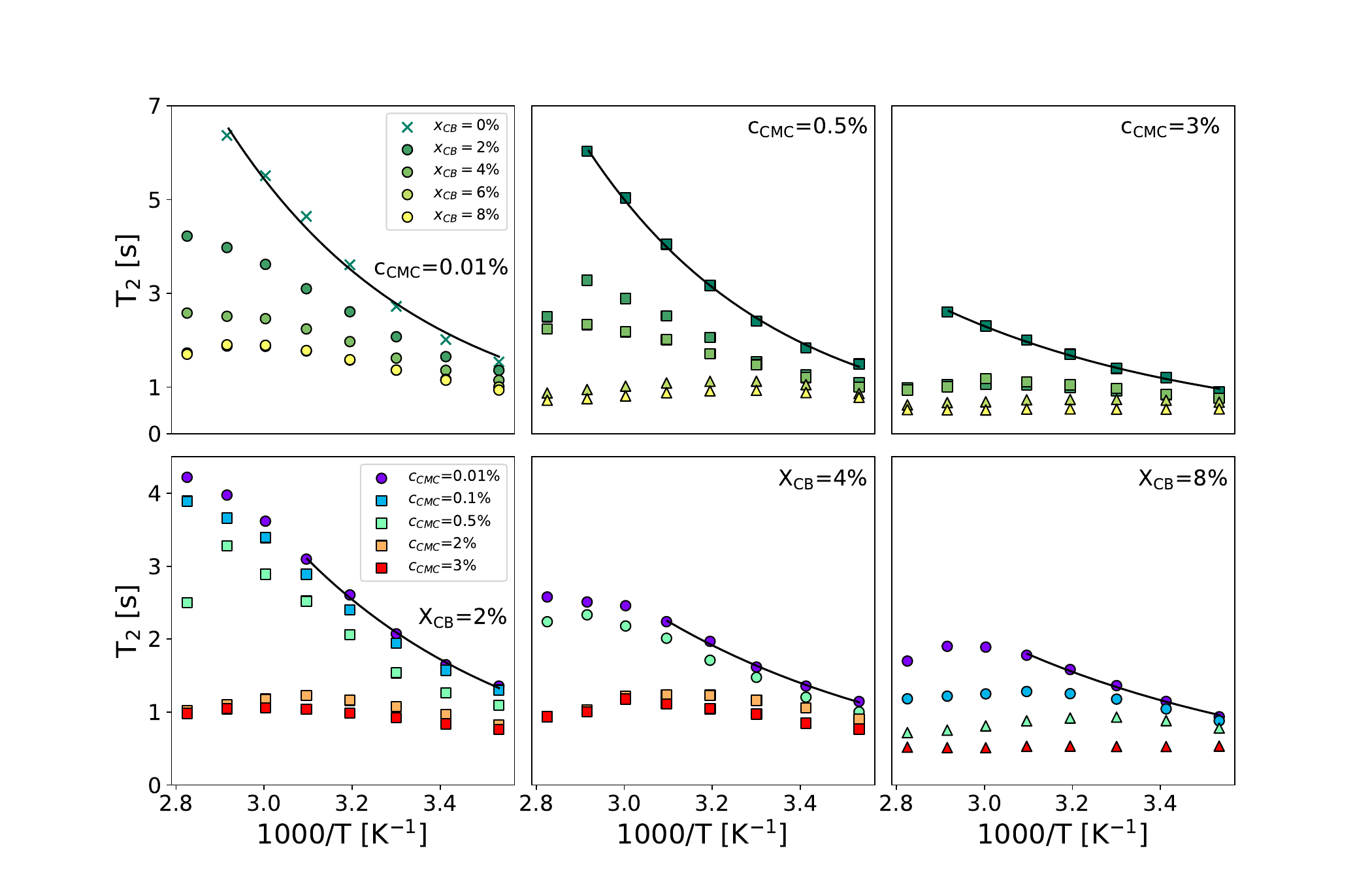}
 \caption{Spin-spin relaxation time $T_2$ vs.~reciprocal temperature for selected formulations varying $x_{\rm CB}$ between 0 and 8 wt.\% for $c_{\rm CMC}=$0.01, 0.5, and 3 wt.\% (top row, left to right) and varying $c_{\rm CMC}$ between 0.01 and 3 wt.\% for $x_{\rm CB}=$2, 4, and 8 wt.\% (bottom row, left to right). Solid lines are Arrhenius fits of data in the whole temperature range (top row) and the 283-323 K range (bottom row). Symbols refer to the rheological behavior: circle and cross:  \textit{viscoelastic liquid}, square: \textit{colloidal gel}, and triangle: \textit{polymer gel}.}
 \label{fgr:figure2}
\end{figure*}

Figure \ref{fgr:figure2} gathers the times $T_2$ measured across all samples investigated between $\rm 10$ and $\rm 80^{\circ}C$. The first row highlights the impact of CB content ($x_{\rm CB}=0-8\ \rm wt.\%$) in three hydrogel matrices made of $0.01$, $0.5$, and $3\ \rm wt.\%$ of CMC, while the second row examines the impact of CMC concentration ($c_{\rm CMC}=0.01-3\ \rm wt.\%$) for three CB contents $2$, $4$, and $8\ \rm wt.\%$. 
At first glance, $T_2$ generally decreases as the reciprocal temperature increases, reflecting the reduced mobility of water molecules. As expected, and consistent with our previous work on CMC hydrogels,\cite{legrand2024acid} $T_2$ in CB-free samples follows an Arrhenius dependence across the entire temperature range, expressed as:
\begin{equation} \label{eq:2}
    T_2=T_2^0 \exp{\left(-E_a/RT\right)}
\end{equation}
where $T_2^0$ is the pre-exponential factor and $E_a$ represents the apparent activation energy that describes the temperature sensitivity of $T_2$ (see solid lines in Fig.~\ref{fgr:figure2}). Although the Arrhenius trend appears to hold at lower temperatures (typically below $50 \rm ^{\circ}C$) in hybrid hydrogels as well, $T_2$ clearly deviates from this behaviour at higher temperatures, indicating more complex dynamic scenarios (discussed further below). In addition, $T_2$ consistently decreases with increasing concentrations of CB and CMC, highlighting that although water molecules are indistinguishable in a given sample, their time-averaged dynamics are strongly influenced by the gel composition. This important finding becomes even more evident when using the alternative representation $T_2^{-1}=f(c_{\rm CMC})$ for various $x_{\rm CB}$ and temperatures, which coincides with the weighted average of the relaxation rates of two fundamental water states, namely \textit{bulk} and \textit{bound}, expressed as 
\begin{equation}\label{eq:3}
    \frac{1}{T_2}=\frac{\phi (T_2^{bulk}-T_2^{bound}) +T_2^{bound}}{T_2^{bulk} T_2^{bound}}
\end{equation}
where $\phi$ is the fraction of bound water molecules assumed to vary as $c_{\rm CMC}$, resulting in $T_2^{bound}\approx 0.05-0.1~\rm s$ and $T_2^{bulk}\approx 1-5~\rm s$ regardless of the gel composition (see \textcolor{blue}{SI Section~1}).\\ 

\begin{figure}[h]
\centering
  \includegraphics[width=1\columnwidth]{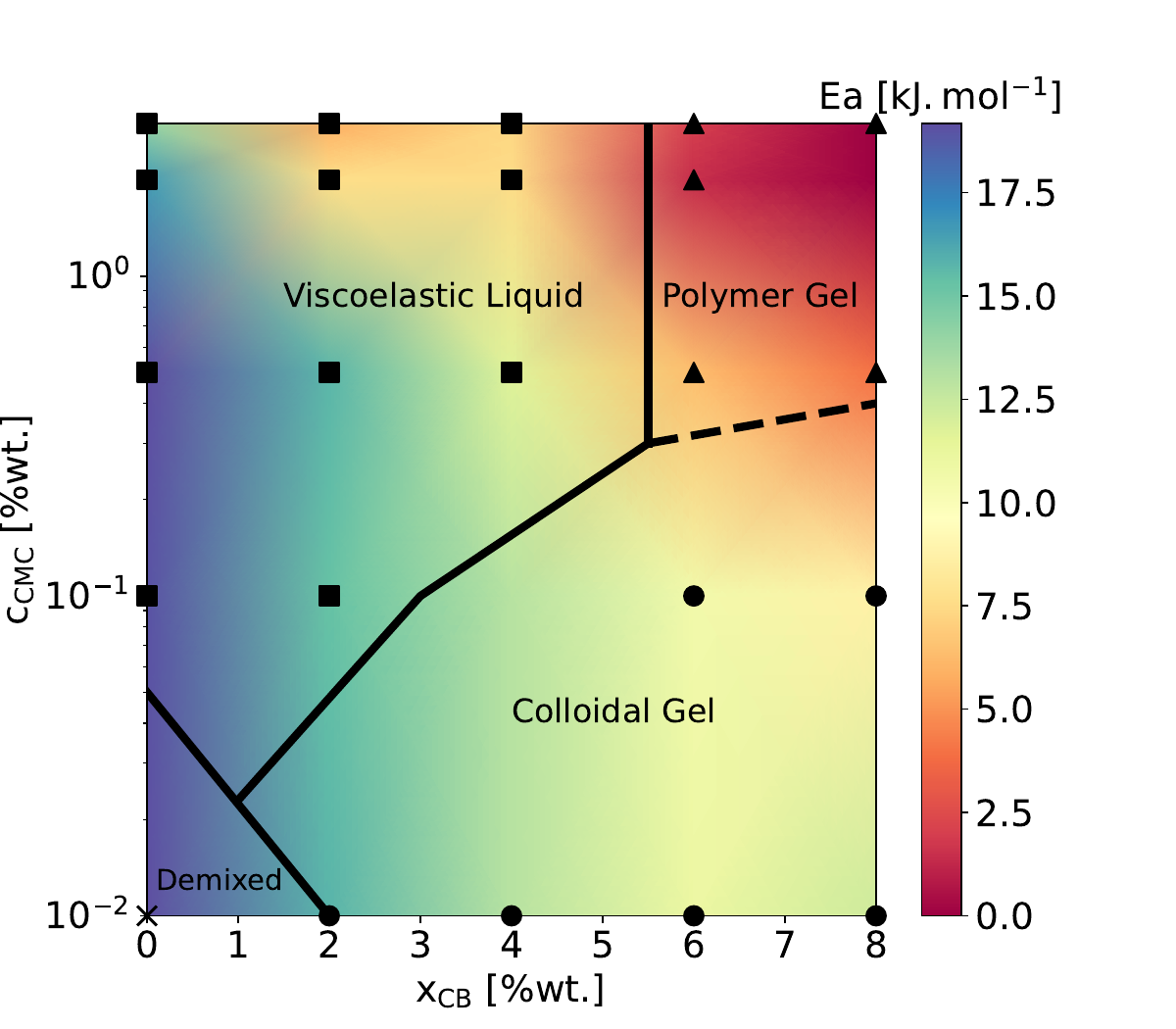}
  \caption{Phase diagram gathering rheological state determined at $T=22^\circ \rm C$ extracted from ref.\cite{legrand2023dual} and low-temperature apparent activation energies reported in Fig.~\ref{fgr:figure2}. Symbols mark the sample formulations investigated in the present work. Squares, triangles, circles, and crosses correspond to viscoelastic liquids, polymer gels, colloidal gels, and a CB-free polymer solution, respectively. The color map is interpolated using the \textit{Gouraud} shading from the \textit{matplotlib} package (Python). $E_a$ values are reported in \textcolor{blue}{SI Section~2}.}
  \label{fgr:diagram}
\end{figure}

Beyond the evolution of $T_2$ with the temperature and the gel formulation, it is also worth noting that $E_a$, extracted from Arrhenius fits between $10^{\circ} \rm C$ and $50^{\circ} \rm C$, clearly decreases as either $c_{\rm CMC}$ or $x_{\rm CB}$ increases. Its values range from ca.~$19~\rm kJ.mol^{-1}$ in the most dilute samples (e.g., CMC0.01-, CMC0.1- and CMC-0.5-CB0), to nearly $0~\rm kJ.mol^{-1}$ in the densest hydrogel (CMC3-CB8), where $T_2$ no longer seems to depend on temperature. Although it might be tempting to link these activation energies to effective binding energies determined in rheology, \cite{baeza2016network} the interpretation in NMR is quite the opposite. Here, a low activation energy corresponds to high thermal stability in a material where water molecules move through a dense network with slow polymer dynamics.\cite{legrand2024rheological} While a quantitative correlation between $E_a$ and the network density/topology remains elusive, comparing these values with rheological properties allows us to build a remarkable phase diagram presented in Figure~\ref{fgr:diagram}. This diagram integrates (i) the sample rheological state -- either \textit{viscoelastic liquid}, \textit{polymer gel}, or \textit{colloidal gel}, as defined above in the introduction (data extracted from ref.\cite{legrand2023dual}) -- and (ii) the $E_a$ values measured from NMR experiments (see the color map). This dual analysis clearly illustrates the above-mentioned trend, showing that $E_a$ decreases with increasing CB or CMC content, and highlights the excellent agreement between the two techniques in defining the \textit{polymer gel} domain, which encompasses the densest materials (red zone). 
In addition, the highest $E_a$ values (purple zone) are only observed in CB-free hydrogels, whereas nearly CMC-free gels display intermediate values (light green) across a wide range of $x_{\rm CB}$, indicating that the impact of CB and CMC on $E_a$ is not equivalent. This observation supports the correlation between macroscopic rheological properties and local water diffusion: a CB-free \textit{liquid}-like sample containing a relatively high fraction of (partly hydrophilic) CMC results in a lower impact on $E_a$ than a \textit{gel} made of a small fraction (e.g., $0.01 \rm wt.\%$) of CMC decorating a physically cross-linked network of (hydrophobic) CB particles. In other words, $T_2$ is more sensitive to the presence of a 3D network (even if loose and made of strands that do not interact with water molecules) than to an increase in the concentration of a polymer that interacts favorably with water in solution. Although not directly comparable, this result aligns with observations in polymer melts, where $T_2$ decreases with increasing cross-link density in a nearly identical chemical environment.\cite{litvinov1998density} The situation becomes more complex when moving away from the left- and bottom-axes of the diagram, where it appears that CMC and CB interact synergistically to influence $E_a$. This translates into the centro-symmetric color gradient originating from the CMC3-CB8 formulation (top-right corner of the diagram). In this region, which includes the densest \textit{viscoelastic liquids} and \textit{colloidal gels}, as well as all the \textit{polymer gels}, $E_a$ values progressively decrease as the material densifies, regardless of whether CB or CMC concentrations are increased. The transition from \textit{colloidal gel} to \textit{viscoelastic liquid} implies, in this case, a decrease of $E_a$ (see e.g., $x_{\rm CB}=4 \rm wt.\%$), suggesting that favorable CMC-water interactions dominate the gelation process.

Beyond the low-temperature regime mainly described by the phase diagram, a closer look at high-temperature data in Figure~\ref{fgr:figure2} reveals a striking non-monotonic evolution of $T_2$ in several formulations, particularly noticeable for $x_{\rm CB}=4\ \rm wt.\%$. Overall, increasing the gel density systematically results in a progressive transition of $T_2(T)$ from (i) a monotonic Arrhenius profile, (ii) a monotonic non-Arrhenius profile, and (iii) a non-monotonic profile (see e.g., the $c_{\rm CMC}=0.01\%$ panel). 
While an Arrhenius profile is commonly expected from water molecules that steadily gain mobility with increasing temperature,\cite{sattig2014nmr} a non-monotonic evolution here suggests a thermally induced modification of the network structure. This is further confirmed by CPMG experiments performed upon cooling, which show a perfect overlap of $T_2$ during \textit{heating} and \textit{cooling} for samples with an Arrhenius profile, while revealing a significant decrease in $T_2$ in the cooling branch for other profiles (see \textcolor{blue}{SI Section~2}). 
Interestingly, this hysteresis becomes more pronounced in samples with higher CB content, suggesting that denser colloid-polymer hydrogels exhibit a stronger tendency to age with temperature changes. Although the low $E_a$ values make it challenging to observe the non-monotonic trend of $T_2$ in dense gels, this observation indicates that the rheological properties of these materials are likely to evolve in a complex manner upon heating. 

In line with this reasoning, Figure~\ref{fgr:rheo} shows the frequency dependence of the storage modulus of CMC3-CB8 measured on three different samples at $10~\rm ^{\circ}C$, $50~\rm ^{\circ}C$, and $70~\rm ^{\circ}C$. All three spectra exhibit a low-frequency plateau, which is linked to the topological density of the gel through entropic elasticity, \cite{treloar1975physics} and a high-frequency power-law evolution with an exponent $\alpha$, corresponding to the relaxation modes of the network strands. As expected, increasing the temperature from $10~\rm ^{\circ}C$ to $50~\rm ^{\circ}C$ accelerates the strand relaxation dynamics and decreases the lifetime of attractive interactions ("stickers"\cite{leibler1991dynamics}), resulting in a significant decrease of $G'$ at both high- and low- frequencies. However, at $70~\rm ^{\circ}C$, the situation is more intriguing. Indeed, at this temperature, the plateau modulus increases significantly (by a factor of $\rm4$), and the power-law exponent $\alpha$ shifts from the Rouse-like value of $0.5$ (observed for $10-50~\rm ^{\circ}C$) to $0.3$, both of which unambiguously indicate a strong densification of the network, in good agreement with the aforementioned NMR results showing a decrease of $T_2$. For completeness, we also confirm that the aging kinetics vary significantly with the temperature (see inset in Fig.~\ref{fgr:rheo}), resulting in a rapid increase of $G'$ at $70~\rm ^{\circ}C$. This observation is in line with the non-monotonic onset of $T_2$ observed in most formulations, from $1000/T \approx 3.0$ (i.e., $T>50~\rm ^{\circ}C$) in Fig.~\ref{fgr:figure2}, which suggests a change in the network microstructure, therefore different aging processes and dynamics.

To summarize, we have demonstrated that low-field NMR experiments that measure the spin-spin relaxation time $T_2$ using the CPMG pulse sequence, are effective in probing the gelation and structural evolution of polymer-colloid hydrogels. Specifically, when using CMC, we observed a single and unstretched exponential decay of the transverse magnetization across all gel formulations and temperatures, indicating that water molecules are indistinguishable at the $T_2$ timescale (approximately $4~\rm s$). For $T<50^\circ \rm C$, $T_2(T)$ allows for extracting an apparent activation energy $E_a$, which can be directly compared to rheological data, offering an alternative perspective to study gelation. In the loosest networks, $E_a$ is about $19~\rm kJ.mol^{-1}$, but it decreases to $0~\rm kJ.mol^{-1}$ in the densest networks, indicating a temperature-independent NMR response. On the other hand, for $T>50^\circ \rm C$, we observed significant network densification and accelerated aging, emphasizing the dominance of CB-CMC interactions over CMC-water and CMC-CMC interactions. These findings are supported by rheological experiments, establishing  NMR relaxometry as a practical and valuable tool for systematically investigating hydrogels from the solvent point of view.

\begin{figure}[t]
\centering
  \includegraphics[width=1\columnwidth]{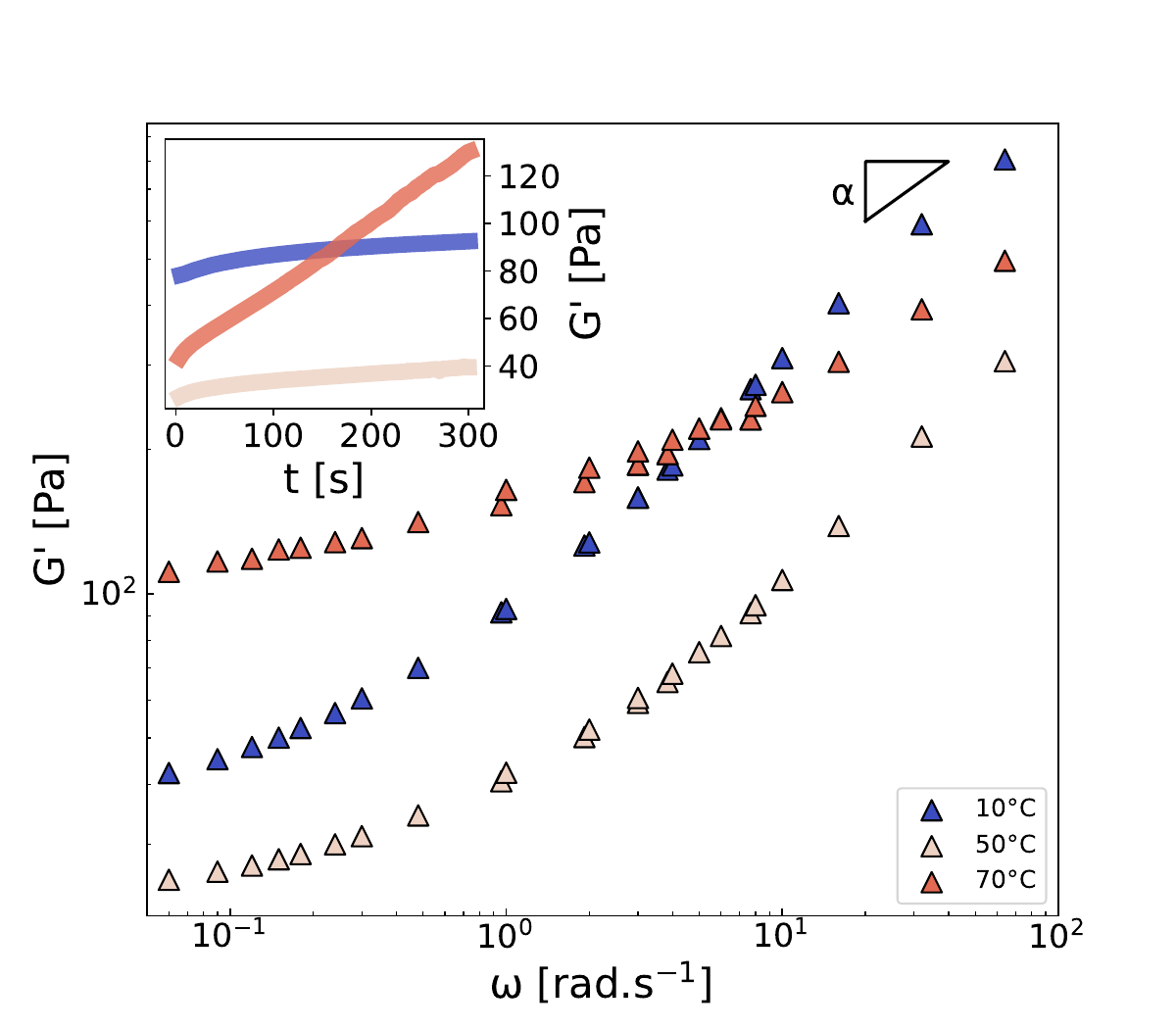}
  \caption{Frequency dependence of the storage modulus $G'$ of the CMC3-CB8 sample measured at 10, 50, and $70~\rm ^{\circ}C$ for $\gamma=0.1$\%. The parameter $\alpha$ is the exponent characterizing the high-frequency power-law response $G'\sim \omega^{\alpha}$. Inset: time sweep measurement performed at $\omega =2\pi~\rm rad.s^{-1}$. See \textcolor{blue}{SI Section~3} for $G''$.}
  \label{fgr:rheo}
\end{figure}

\section*{Experimental Section}
\subsection*{Samples preparation}
\label{subsec:sample_prep}
Samples are prepared by first dissolving sodium carboxymethylcellulose (NaCMC, Sigma Aldrich, $M_w=250~\rm kg.mol^{-1}$ and $\mathrm{DS}=0.9$ as specified by the manufacturer, (see ref.~\cite{legrand2024rheological}for actual measured values) in deionized water (H$_2$O). Stock solutions up to 5\% wt.~are prepared and stirred at room temperature for 48~h until homogeneous, before adding the Carbon Black (CB) particles (VXC72R, Cabot). Samples are placed in a sonicator bath for two rounds of $90$~min separated by a period of 24~h under mechanical stirring. The samples are finally left at rest for another 24~h before being tested. The samples are characterized by their CMC concentration $c_{\rm CMC}$ expressed in $\rm wt.\%$ of the stock solution (without CB) and their mass fraction of CB $x_{\rm CB}$, following \cite{Legrand:2023}.

\subsection*{LF-NMR}
\label{subsec:LF_NMR}
The $\rm ^1H$ low-field NMR experiments were performed on a Bruker minispec mq20 spectrometer (‘‘NF’’ electronics) operating at a resonance frequency of $20~\rm MHz$ with $\rm 90^{\circ}$ and $\rm 180^{\circ}$ pulse lengths of $\rm 2.2\ ms$ and $\rm 4.8\ ms$ respectively, and a dead time of $\rm 15\ \mu s$. A Car-Purcell-Meiboom-Gill (CPMG) echo train acquisition was used to measure data up to about three times the spin-spin relaxation time $T_2$ of gels, being approximately $1~\rm s$. The echo-time was systematically set to $2\tau=2~\rm ms$. It was varied down to $2\tau=0.25\ \rm ms$ for a single sample (CMC2-CB2) at $\rm 10^{\circ}C$ and $\rm 80^{\circ}C$ showing no and little effect on the measurements of $T_2$, respectively (see \textcolor{blue}{SI Section~4}). Measurements were performed by increasing the temperature in a step-wise manner from $\rm 10^{\circ}C$ to $\rm 80^{\circ}C$ every $\rm 10^{\circ}C$ with a BVT 3000 heater connected to a liquid nitrogen Dewar. Before each measurement, the temperature was left to stabilize for $10~\rm min$. For some samples, the experiments were then repeated from high to low temperatures to investigate aging effects (see \textcolor{blue}{SI Section~2}).

\subsection*{Rheology}
\label{subsec:rheometry}
The rheological properties were measured with a cone-and-plate geometry (angle 2$^\circ$, radius $20~\rm mm$ and truncation $46~\rm \mu$m) connected to a strain-controlled rheometer (ARES G2, TA Instruments). The cone was smooth and the plate was sandblasted with a surface roughness of about 1~$\mu$m to prevent wall slip. For a given sample, each temperature was probed independently with a different loading and independent gap calibration. The three temperatures probed $T = 283, 323,$ and $343$~K were maintained constant by a Peltier modulus placed under the bottom plate. The rheological protocol for linear viscoelastic characterization of the samples was divided into three consecutive steps: ($i$) a preshear at $\dot \gamma=500$~s$^{-1}$ for $3~\rm min$ to erase the loading history and rejuvenate the sample \citep{Viasnoff:2002,Bonn:2017,Joshi:2018}; ($ii$) a recovery phase of $5~\rm min$ during which we monitored the sample linear viscoelastic properties by applying small amplitude oscillations in the linear regime at a frequency $\omega = 2\pi ~\rm rad.s^{-1}$; ($iii$) a frequency sweep performed by mutliwave strain signals \cite{Mours:1994} in the linear regime to span frequencies from $\omega = 0.15 $ to $420 ~\rm rad.s^{-1}$. The samples remained in the rheometer for 9~minutes, and no evidence of alternation due to evaporation was witnessed by visual inspection, even at the highest temperature.

\section*{Author contributions}
CRediT: \textbf{Léo Hervéou} conceptualization, data curation,
formal analysis, investigation, methodology, software, visualization,
writing-review \& editing; \textbf{Gauthier Legrand}
conceptualization, investigation, validation, writing-review \& editing; 
\textbf{Thibaut Divoux} conceptualization, methodology, project administration, resources, supervision, visualization, validation, writing-original draft,  writing-review \& editing; 
\textbf{Guilhem P. Baeza} conceptualization, methodology, project administration, resources, supervision, visualization, validation, writing-original draft,  writing-review \& editing;

\section*{Conflicts of interest}
There are no conflicts to declare.

\section*{Data Availability}
Data for this article, including CPMG measurements $I(t)$ for all the samples are available \href{https://www.researchgate.net/publication/384054735_Herveou_et_al_CPMG}{here}. DOI: 10.13140/RG.2.2.30502.33604

\section*{Acknowledgements}
All the authors warmly thank Carlos Fernandez-de-Albá (IMP, INSA-Lyon) and Cédric Lorthioir (LCMCP, Paris) for technical support and insightful discussions about low-field NMR experiments. This work was supported by the LABEX iMUST of the University of Lyon (ANR-10-LABX-0064), created within the ``Plan France 2030" set up by the French government and managed by the French National Research Agency (ANR).

\providecommand*{\mcitethebibliography}{\thebibliography}
\csname @ifundefined\endcsname{endmcitethebibliography}
{\let\endmcitethebibliography\endthebibliography}{}

\newpage

\onecolumn
\setcounter{equation}{0}
\setcounter{figure}{0}
\global\def\thefigure{S\arabic{figure}}
\setcounter{table}{0}
\global\def\thetable{S\arabic{table}}

\begin{center}
  {\LARGE{Supporting Information}}
\end{center}

\section*{1. Impact of the CB content on $\rm T_2$}

\begin{figure}[!h]
\centering
  \includegraphics[width=1.1\columnwidth]{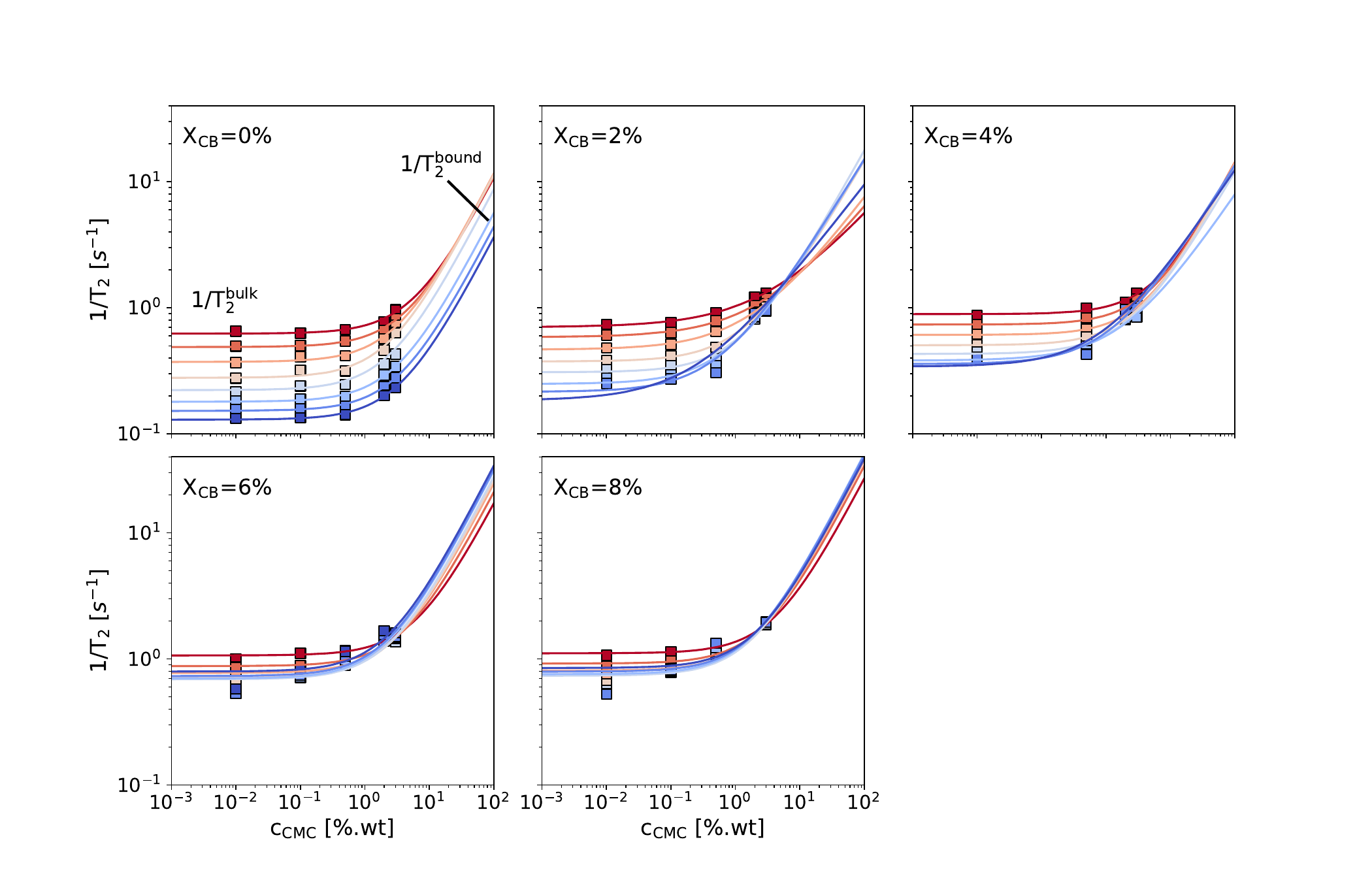}
  \caption{Spin-spin relaxation rate $1/T_2$ vs.~CMC concentration $c_{\rm CMC}$ measured various temperatures, every $10~\rm ^{\circ}C$ between $10~\rm ^{\circ}C$ and $80~\rm ^{\circ}C$ (see color code from dark blue to red) for the pure CMC solution, and hybrid hydrogels containing $x_{\rm CB}=$2, 4, 6, and 8 wt.\% of carbon black particles. Solid lines show the best fit to the data with Eq.(3) from the main text.}
  \label{fgr:formulation}
\end{figure}

Figure~\ref{fgr:formulation} shows the impact of the CMC concentration on the transverse relaxation rate $1/T_2$ for hydrogels containing 0, 2, 4, 6 and 8 $\rm wt.\%$ in CB. Data were measured at various temperatures, every $10^\circ \rm C$ between $\rm 10$ to $\rm 80^{\circ}C$. Following Eq.~(3) in the main text, the low and high $\rm c_{cmc}$ limit values represent $1/T_2^{bulk}$ and $1/T_2^{bound}$. For each CB content, it is worth noting that while $1/T_2^{bulk}$ tends to increase upon increasing the temperature (corresponding to longer $T_2$), the extrapolated values of $1/T_2^{bound}$ mostly collapse in a narrow interval close to $1/T_2^{bound}\approx 10~\rm s^{-1}$ corresponding to $T_2^{bound}\ll 4 ~\rm s$, in agreement with a strongly reduced molecular mobility of the water molecules. This dichotomy is further reminiscent of the two mechanisms impacting the $T_2$ values evoked in the main text: i) the presence of a 3D network limiting long-range water diffusion (mostly observed at low CB and CMC contents), and ii) short-range attractive interactions between the CMC and the water molecules (dominating $T_2$ at large CMC content).  

\clearpage

\section*{2. Heating vs.~Cooling $\rm T_2$ measurements}

\begin{figure}[h]
\centering
  \includegraphics[width=0.8\columnwidth]{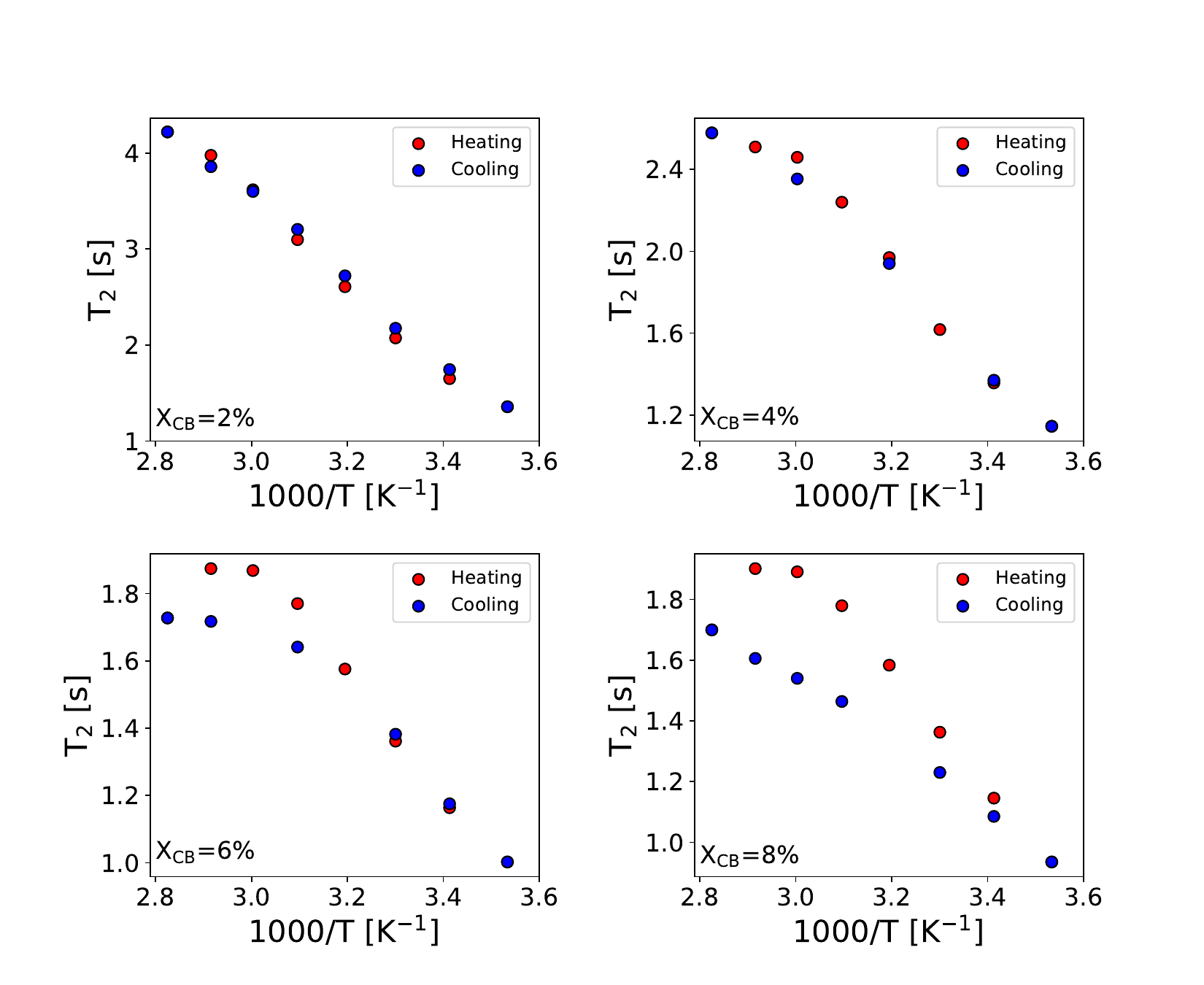}
  \caption{Spin-spin relaxation time $T_2$ vs.~reciprocal temperature measured upon heating from $\rm 10$ to $\rm 80^{\circ}C$ and subsequent cooling over the same temperature range. Experiments performed on selected formulations where $c_{\rm CMC}=0.01~\rm wt.\%$ and $x_{\rm CB}=$2, 4, 6, and 8 wt.\%.}
  \label{fgr:hysteresis}
\end{figure}

\begin{figure}[h]
\centering
  \includegraphics[width=0.8\columnwidth]{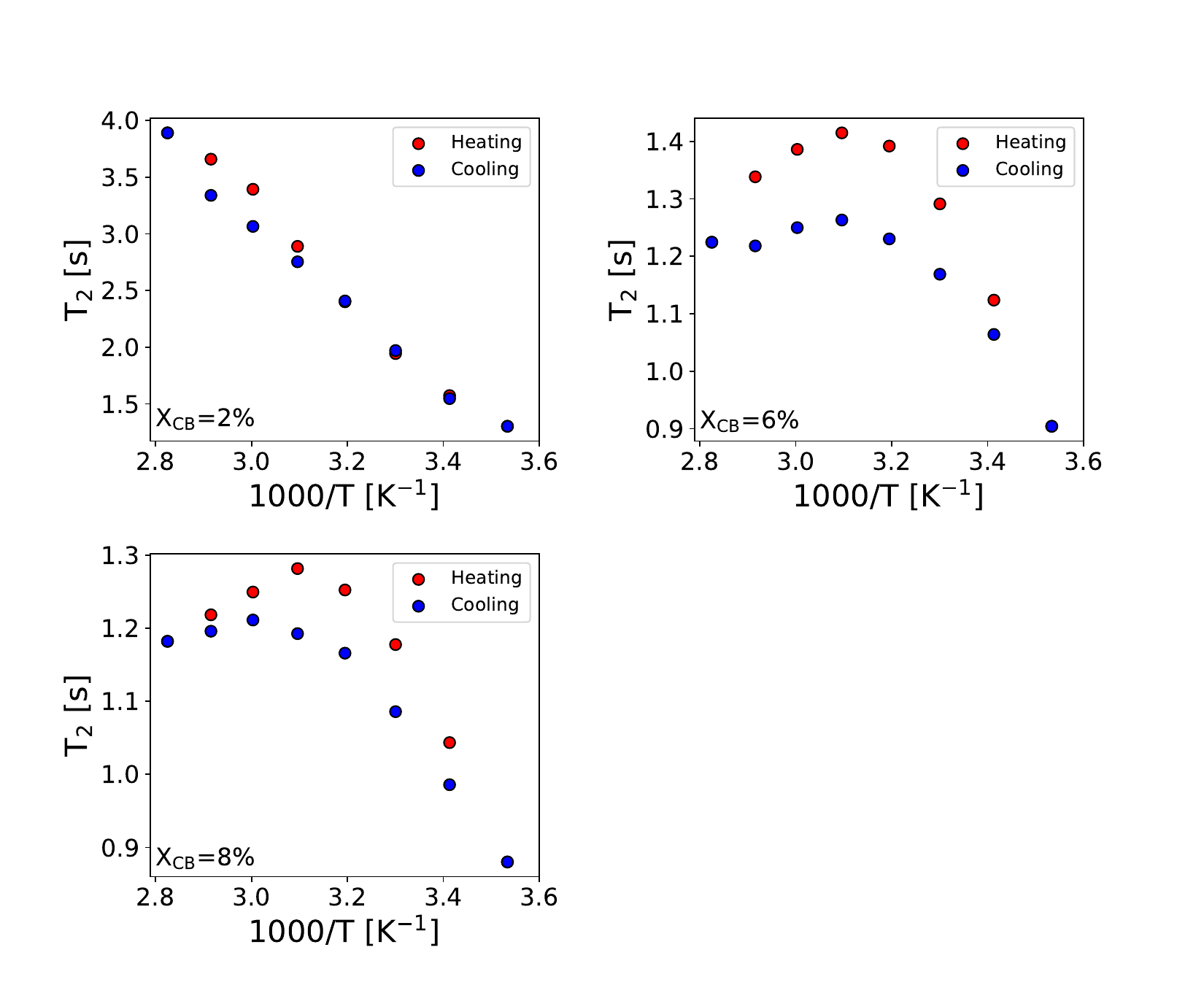}
  \caption{Spin-spin relaxation time $T_2$ vs.~reciprocal temperature measured upon heating from $\rm 10$ to $\rm 80^{\circ}C$ and subsequent cooling over the same temperature range. Experiments performed on selected formulations where $c_{\rm CMC}=0.1$wt.\% and $x_{\rm CB}=$2, 6 and 8 wt.\%.}
  \label{fgr:hysteresis2}
\end{figure}

Figure~\ref{fgr:hysteresis} shows the impact of the thermal history on $T_2$ measured on samples containing $c_{\rm CMC}=0.01~\rm wt. \%$ and various fractions of CB. The measurements are performed in a stepwise manner by increasing the temperature from $\rm 10$ to $\rm 80^{\circ}C$ and subsequently cooling down from $\rm 80$ to $\rm 10^{\circ}C$. While no significant effect is observed at $x_{\rm CB}=2~\rm wt.\%$, the data on the \textit{cooling} branch systematically exhibit lower $T_2$ values corresponding to a growing degree of aging when increasing the gel density. Figure~\ref{fgr:hysteresis2} confirms this result by showing even stronger effects for $c_{\rm CMC}=0.1~\rm wt.\%$. 

\clearpage

\section*{3. Loss modulus and loss factor of CMC3-CB8 at various temperatures}

Figure~\ref{fgr:rheo} shows the loss modulus, $G''$, and loss factor, $\tan \delta =G''/G'$, of the CMC3-CB8 hydrogel corresponding to the storage modulus $G'$ reported in Figure 4 in the main text. These data confirm the non-monotonic behavior of the viscoelastic properties upon increasing the temperature, particularly emphasizing the relative decrease of its viscous character at $70^{\circ}\rm C$.

\begin{figure}[h]
\centering
  \includegraphics[width=.6\columnwidth]{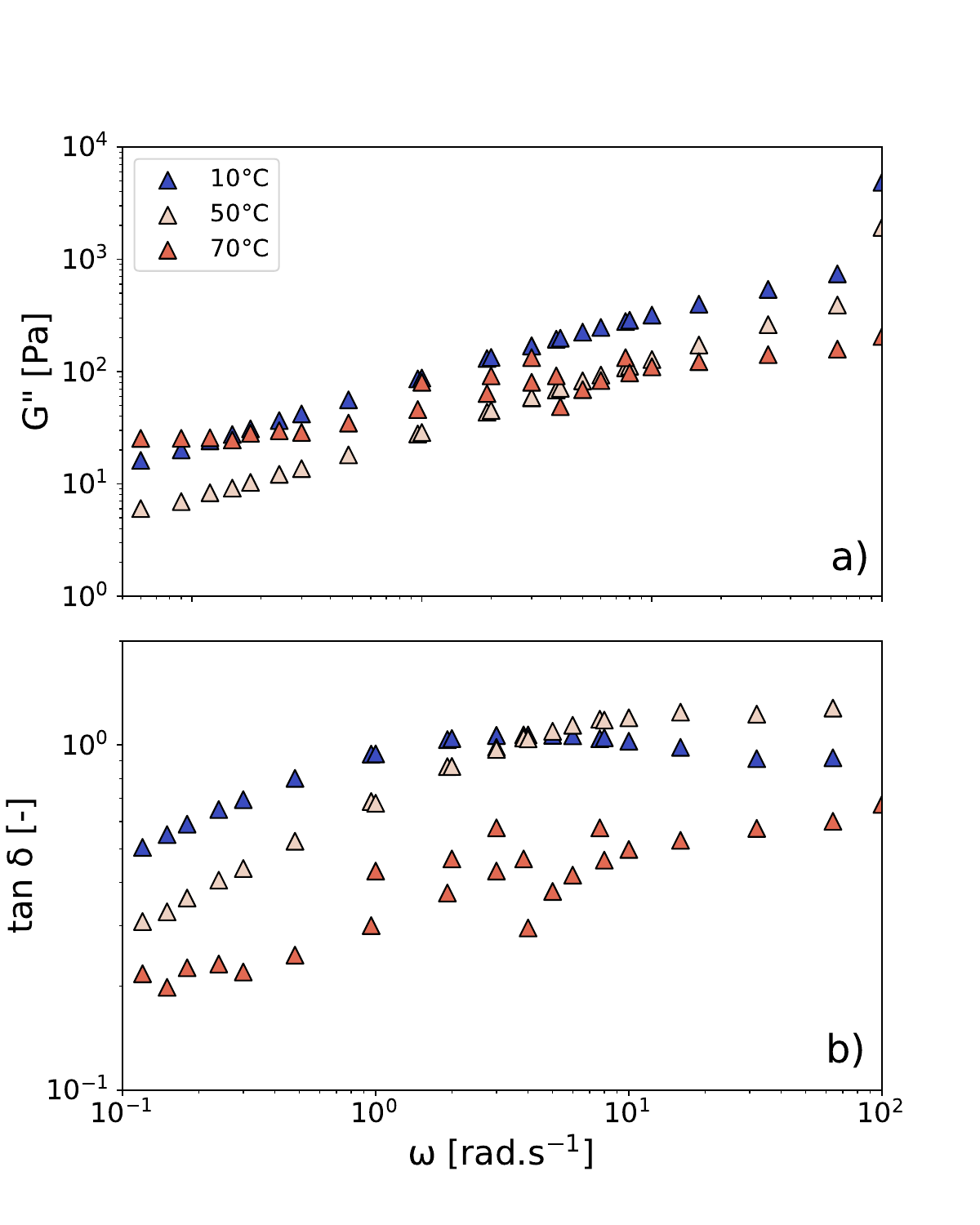}
  \caption{Frequency dependence of (a) the loss modulus $G''$ and (b) loss factor $\tan \delta =G''/G'$ of the CMC3-CB8 sample measured at 10, 50, and $70~\rm ^{\circ}C$ for $\gamma=0.1$\%.}
  \label{fgr:rheo}
\end{figure}

\clearpage

\section*{4. Impact of the CPMG echo time on the $\rm T_2$ values at low and high temperatures.}

\begin{figure}[h]
\centering
  \includegraphics[width=0.7\columnwidth]{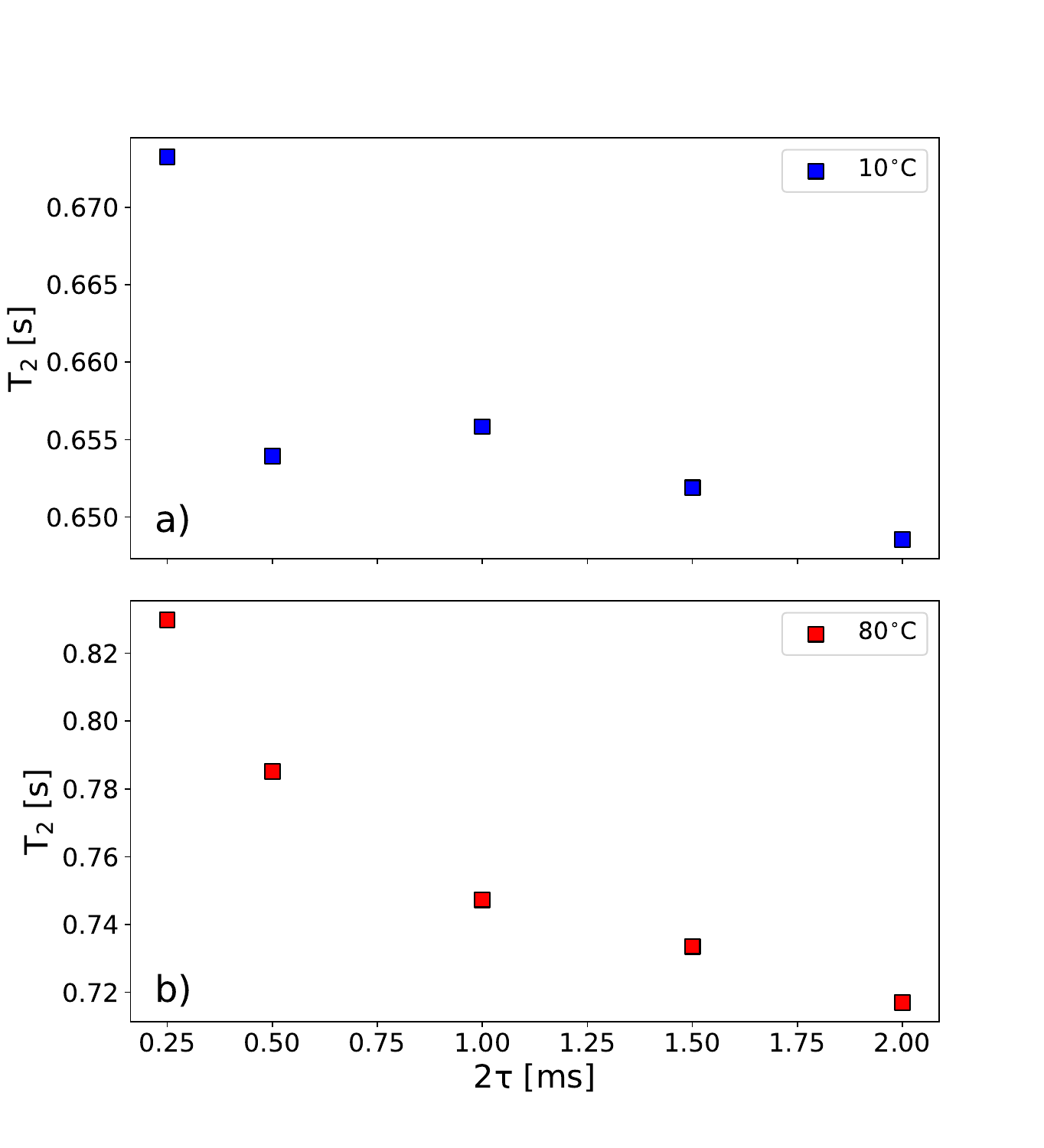}
  \caption{Spin-spin relaxation time $T_2$ vs.~CPMG echo-time $2\tau$ for CMC2-CB2 measured at (a) $\rm 10^{\circ}C$ and (b) $\rm 80^{\circ}C$ on independent samples.}
  \label{fgr:echo}
\end{figure}

Figure~\ref{fgr:echo} shows $T_2$ values obtained for the sample CMC2-CB2 at $\rm 10^{\circ}C$ and $\rm 80^{\circ}C$ across different  CPMG echo-times ($2\tau$). Although a slight decrease in $T_2$ is observed with increasing echo-time in both cases, it only corresponds to a relative variation of about $\rm -4\%$ and $\rm -16\%$, respectively. These variations are negligible compared to the potential aging effects and the significant $T_2$ changes with temperature reported in Figure~2 of the main text. This key observation confirms that our CPMG measurements are not impacted by technical artefacts related to the nature of the material.\cite{jarenwattananon2018breakdown}

\end{document}